\documentclass[12pt]{article}
\usepackage{amsmath,amsfonts,amssymb,cite}
\usepackage[english]{babel}
\let\vph\varphi

\newcommand{\sn}{\,\text{sn}\,\!}
\newcommand{\cn}{\,\text{cn}\,\!}
\newcommand{\dn}{\,\text{dn}\,\!}

\let\p\partial

\def\bq{ \begin{equation} }
\def\eq{ \end{equation} }
\def\ben{ \begin{eqnarray} }
\def\en{ \end{eqnarray} }

\newcounter{theo}
\newcommand{\theo}{\addtocounter{theo}{1}\textbf{Theorem \thetheo.} }
\newcounter{lem}

\newcounter{rem}

\newcounter{defi}

\numberwithin{equation}{section}

\begin{document}

\baselineskip=15pt
\vspace{1cm} \centerline{{\LARGE \textbf {Integrable hyperbolic equations
 }}}
 \vspace{0.5cm} \centerline{{\LARGE \textbf {of sin-Gordon type.
 }}}

\vskip1cm \hfill
\begin{minipage}{13.5cm}
\baselineskip=15pt {\bf A.G. Meshkov${}^{1}$,   V.V. Sokolov${}^{2}$}
\\ [2ex] {\footnotesize
${}^{1}$ \ Oryol State Technical University
\\
${}^{2}$ \ Landau Institute for Theoretical Physics
}\\
\mbox{}\hspace{1.4ex} \textbf{E-mail}:  sokolov@itp.ac.ru
\end{minipage}


\begin{abstract}
A complete list of nonlinear one-field hyperbolic equations having generalized integrable x- and y-symmetries of the third order is presented. 
The list includes both sin-Gordon type equations and equations linearizable by differential substitutions.
\end{abstract}
Keywords: generalized symmetry,   sin-Gordon type equation, Liouville type equation.\\[2mm]
MSC: 37K10, 35Q53, 35Q58
\thispagestyle{empty}
\baselineskip6.5mm

\section{Introduction}

The symmetry approach to classification of integrable PDEs (see surveys \cite{ss,umn,mss} and references there) is
based on the existence of higher infinitesimal symmetries and/or
conservation laws for integrable equations. This approach especially efficient for
evolution equations with one spatial variable. In particular, all integrable
equations of the form
\begin{equation}\label{kdv}
u_t=u_3+F(u_2,\, u_1, \, u ), \qquad u_{i}=\frac{\partial^{i} u}{\partial x^{i}}
\end{equation}
were found in \cite{soksvin1,soksvin2}. The following  list of integrable
equations

{\bf List 1}:
\begin{align}
&u_t=u_{xxx}+uu_x,\ \label{list1}\\
&u_t=u_{xxx}+u^2u_x, \label{list2}\\
&u_t=u_{xxx}+u_x^2, \label{list3}\\
&u_t=u_{xxx}-\frac{1}{2}u_x^3+(c_1e^{2u}+c_2e^{-2u})u_x, \label{list4}\\
&u_t=u_{xxx}-\frac{3u_xu_{xx}^2}{2(u_x^2+1)}+c_1(u_x^2+1)^{3/2}+c_2u_x^3, \label{list5}\\
&u_t=u_{xxx}-\frac{3u_xu_{xx}^2}{2(u_x^2+1)}-\frac{3}{2} \wp(u)u_x(u_x^2+1), \label{list6}
\end{align}
\begin{align}
&u_t=u_{xxx}-\frac{3u_{xx}^2}{2u_x}+\frac{3}{2 u_x}-\frac{3}{2} \wp(u)u_x^3, \label{list7}\\
&u_t=u_{xxx}-\frac{3u_{xx}^2}{2u_x}, \label{list8}\\
&u_t=u_{xxx}-\frac{3u_{xx}^2}{4u_x}+c_1u_x^{3/2}+c_2u_x^2,\ \ c_1\ne0\ \text{or } c_2\ne0, \label{list9}\\
&u_t=u_{xxx}-\frac{3u_{xx}^2}{4u_x}+cu,  \label{list10}\\
&\begin{aligned} \label{list11}
u_t&=u_{xxx}-\frac34\,\frac{u_{xx}^{2}}{{u_{x}}+1}+3\,u_{xx}u^{-1}(\sqrt{u_{x}+1}-u_{x}-1)\\
&-6 \,u^{-2} u_{x}(u_{x}+1)^{3/2}+3 \,u^{-2} u_{x}\,(u_{x}+1)(u_{x}+2) ,
\end{aligned}
\end{align}
\begin{align}
&\begin{aligned} \label{list12}
u_t&=u_{xxx}-\frac34\,\frac {u_{xx}^{2}}{{u_{x}}+1}-3\,\frac{u_{xx}\,(u_{x}+1)\cosh {u}}{\sinh{u}}
+3\,\frac {u_{xx}\sqrt {u_{x}+1}}{\sinh{u}}\\
&-6\,\frac {u_{x}(u_{x}+1)^{3/2}\cosh{u}}{\sinh^{2}u}
+3\,\frac {u_{x}\,(u_{x}+1)(u_{x}+2)}{\sinh^{2}u}+u_{x}^2(u_{x}+3),
\end{aligned} \\
&u_t=u_{xxx}+3u^2u_{xx}+3u^4u_x+9uu_x^2,  \label{list13}\\
&u_t=u_{xxx}+3uu_{xx}+3u^2u_x+3u_x^2, \label{list14}\\
&u_t=u_{xxx}. \label{list14a}
\end{align}
can be derived from  \cite{soksvin1}. Here $ ( \wp')^2=4 \wp^3-g_2 \wp-g_3$, and $k,c,c_1,c_2, g_2, g_3$ are arbitrary constants.
Equations
(\ref{list1})--(\ref{list9}) are integrable by the inverse scattering method whereas (\ref{list10})--(\ref{list14}) are linearizable
($S$ and $C$-integrable in the terminology by F. Calogero). The above list is complete up to transformations of the form
\begin{equation}\label{tr1}
u\to \phi(u);\ \ t\to t,\ x\to x+c t;\ \ x\to\alpha x,\, \ t\to\beta t, \,\ u\to \lambda u; \ \ u\to u+\gamma x +\delta t.
\end{equation}
 The latter transformation  preserves the form (\ref{kdv}) only for equations with  $\frac{\partial F}{\partial u}=0$. Moreover
the linear equations   admit the transformation:
\begin{equation}\label{tr1a}
u\to u\exp(\alpha  x+\beta  t).
\end{equation}

Since the symmetry approach is pure algebraic, the function $\phi$ and
the constants $c,\alpha, \beta,\lambda , \gamma$ and $\delta$ supposed to be complex-valued.
Thus, we do not distinguish equations
$u_{t}= u_{xxx}-u_{x}^{3}$ and $u_{t}= u_{xxx}+u_{x}^{3}$ and so on.

For scalar hyperbolic equations of the form
\begin{equation}\label{hyper}
u_{xy}= \Psi(u, u_{x}, u_{y})
\end{equation}
the symmetry approach postulates the existence of both $x$-symmetries
\begin{equation}\label{sym1}
u_{t}=A(u, u_{x}, u_{xx}, \dots, ),
\end{equation}
and $y$-symmetries
\begin{equation}\label{symm2}
u_{t}=B(u, u_{y}, u_{yy}, \dots, ).
\end{equation}
Two equations (\ref{hyper}) are called {\it equivalent} if they are related by transformations of the form
\begin{align}
&x\leftrightarrow y;\ \
u\to \phi(u);\ \  x\to\alpha x,\, \ y\to\beta y, \,\ u\to \lambda u; \ \ u\to u+\gamma x +\delta y, \label{ptr}\\
\intertext{{  and for the linear equations by}}
&u\to u\exp(\alpha  x+\beta y);\qquad u\to u+c\, x\,y. \label{ptrL}
\end{align}
 Here, in general,  the function $\phi$ and the constants supposed to be complex-valued.

For the famous integrable $\sin$-Gordon\footnote{We do not
distinguish $\sin$-Gordon and $\sinh$-Gordon equations} equation
\begin{equation}\label{sin}
u_{xy}=c_1 e^u+c_2 e^{-u}
\end{equation}
the simplest $x$ and $y$-symmetries are given by
$$
u_{t}= u_{xxx}-\frac{1}{2} u_{x}^{3},\qquad u_{t}= u_{yyy}-\frac{1}{2}
u_{y}^{3}.
$$
These evolution equations are integrable themselves (a special case of equation (\ref{list4})).

The general higher symmetry classification for equations (\ref{hyper}) turns out to be very complicated problem,
which is not solved till now. Some important special results have been obtained in  \cite{zibshab1,zibshab2,zibsok}.
In general, all three functions $\Psi, A,B$ should be found from the compatibility
conditions for equations (\ref{hyper}), (\ref{sym1}) and (\ref{hyper}),
(\ref{symm2}). However, if
the functions $A$ and $B$ are somehow fixed, then it is not difficult to
verify whether the corresponding function $\Psi$ exists or not and to find it.

To describe all integrable equations (\ref{hyper}) of the sin-Gordon type, we
assume (see Discussion) that both symmetries (\ref{sym1}) and (\ref{symm2}) are  {\it integrable} equations of the form
\begin{equation} \label{kdvtype}
u_{t}= u_{xxx}+F(u,u_{x},u_{xx}),\qquad u_{t}= u_{yyy}+G(u,u_{y},u_{yy}).
\end{equation}
It turns out that taken for $x$-symmetry one of the equations of the List 1, one can easily find the corresponding
equations (\ref{hyper}) having $y$-symmetries or can  prove that such equations do not exist. In the
Section 2 we present all integrable hyperbolic equations thus obtained.

The hyperbolic equations can be separated by  presence or absence of $x$
and $y$-integrals (see Discussion). Consider, for instance, the Liouville equation
$$
u_{xy}= e^u.
$$
It is easy to verify that the function
$$
  P=u_{xx}-\frac 12u_x^2
$$
does not depend on $y$ (i.e. is a function depending on $x$ only)  for any solution $u(x,y)$ of the Liouville equation.
Analogously, the function
$$
  Q=u_{yy}-\frac 12u_y^2
$$
does not depend on $x$ on the solutions of the Liouville equation.

A function $w(x,y,u,u_y,u_{yy},\ldots)$ that does not depend on $x$
for any solution of (\ref{hyper}) is called $x$-{\it integral}.
Similarly  the $y$-integrals are defined. An equation of the form
(\ref{hyper}) is called {equation of the Liouville type} (or Darboux integrable
equation), if the equation possesses both nontrivial $x$- and $y$-integrals.
Some of integrable hyperbolic equations found in Section 2 are equations
of the  Liouville type.

 In contrast, the sin-Gordon equation (\ref{sin}) has no $x$-
and $y$-integrals for generic values of the constants $c_{i}$. There are two types of such equations. 
Equations of the first type can be reduced to the linear Klein-Gordon  equation $u_{xy}=c u$ by differential substitutions. 
If an equation with the third order symmetries has no integrals and linearizing substitutions, we call it {\it equation of sin-Gordon type}. 
The following equations from the lists of Section 2 are equations of this kind:
\begin{align}
u_{xy}&=c_1e^u+c_2e^{-u}; \label{sh2v}\\
u_{xy}&= f(u)\sqrt{\vphantom{u_y^2}u_x^2+1}, \ \ \ f''=c f; \label{sh1v}\\
u_{xy}&= \sqrt{\vphantom{u_y^2}u_x}\sqrt{u_y^2+1}; \label{sh3v}\\
u_{xy}&= \sqrt{ \wp(u)-\mu}\sqrt{\vphantom{u_y^2}u_x^2+1}\sqrt{u_y^2+1}. \label{sh4v}
\end{align}
Here $( \wp')^2=4 \wp^3-g_2 \wp-g_3,\,4 \mu^3-g_2 \mu-g_3=0$ and $c,c_1,c_2,a,\mu,g_2,g_3$ are
constants. Equations (\ref{sh1v}), (\ref{sh4v}) are known. These equations are related to equation (\ref{sh2v}) 
via differential substitutions \cite{zibsok,Bor}.  
Equation (\ref{sh3v}) is probably new. The corresponding differential substitution is given by
$$
v_{xy}=\frac12 \cosh v, \qquad  u_{xy}=\sqrt{ u_x}\sqrt{u_y^2+1}, \qquad
v=\ln(u_y+\sqrt{u_y^2+1}).
$$


\section{Hyperbolic  equations with third order symmetries}

\theo {\it Suppose both $x$- and $y$-symmetry of a hyperbolic equation of the form} (\ref{hyper})
{\it belong to the list} (\ref{list1})--(\ref{list14a}) {\it up to transformations} (\ref{tr1}), (\ref{tr1a}). 
{\it Then
this equation  belongs to the following list:
\begin{align}
u_{xy}&= f(u)\sqrt{\vphantom{u_y^2}u_x^2+1}, \ \ \ f''=c f, \label{sh1}\\
u_{xy}&=ae^u+be^{-u}, \label{sh2}\\
u_{xy}&= \sqrt{\vphantom{u_y^2}u_x}\sqrt{u_y^2+1}, \label{sh3}\\
u_{xy}&= \sqrt{\vphantom{u_y^2}u_x^2+1}\sqrt{u_y^2+1}, \label{sh4a}\\
u_{xy}&= \sqrt{\wp(u)-\mu}\sqrt{\vphantom{u_y^2}u_x^2+1}\sqrt{u_y^2+a}, \label{sh4}\\
u_{xy}&=2uu_x,   \label{sh6}  \\
u_{xy}&=2u_x\sqrt{u_y},  \label{sh7}\\
u_{xy}&=u_x\sqrt{u_y^2+1}. \label{sh8}\\
u_{xy}&= \sqrt{\rule{0pt}{1.4ex}u_xu_y}, \label{sh10}\\
u_{xy}&= \frac{u_x(u_y+a)}{u},\ \ a\ne0, \label{sh10a}\\
u_{xy}&=(ae^u+be^{-u})u_x, \label{sh10c} \\
u_{xy}&= u_y\eta\,\sinh^{-1} u \big(\eta\,e^{u}-1\big),  \label{sh11a}\\
u_{xy}&=\frac{2u_y\eta }{\sinh u}(\eta \cosh u-1),  \label{sh11c}\\
u_{xy}&= \frac{2\xi \eta }{\sinh u }\big((\xi \eta +1)\cosh u-\xi -\eta \big),   \label{sh11b}\\
u_{xy}&= u^{-1}u_y\,\eta\,(\eta-1)+c\,u\,\eta\,(\eta+1), \label{sh14b}\\
u_{xy}&= 2u^{-1}u_y\,\eta\,(\eta -1), \label{sh14a}\\
u_{xy}&= 2u^{-1}\xi\,\eta\,(\xi-1)(\eta-1 ), \label{sh14}\\
u_{xy}&=u^{-1}u_xu_y-2u^2u_y,\label{sh12}\\
u_{xy}&=u^{-1}u_x(u_y+a)-uu_y  \label{sh13}\\
u_{xy}&=\sqrt{u_y}+au_y,  \label{shL}\\
u_{xy}&=cu,  \label{sh10b}
\end{align}
up to transformations} (\ref{ptr}), (\ref{ptrL}). {\it Here $\wp$ is the Weierstrass function:}
$( \wp')^2=4 \wp^3-g_2 \wp-g_3,\quad$ $4 \mu^3-g_2 \mu-g_3=0,\quad$
$ \xi =\sqrt{u_y+1},\quad \eta =\sqrt{u_x+1}$; \ $a,b,c,\mu,g_2,g_3$
{\it are constants.}

{\bf Proof.}  If (\ref{kdvtype}) is an $x$-symmetry for (\ref{hyper}),
then
\begin{equation}\label{opreq1}
\frac{d^{2}}{d x dy}(u_{xxx}+F)=\frac{\partial \Psi}{\partial u_{x}} \frac{d}{d
x}(u_{xxx}+F)+\frac{\partial \Psi}{\partial u_{y}} \frac{d}{dy}(u_{xxx}+F)+\frac{\partial \Psi}{\partial u}(u_{xxx}+F).
\end{equation}
Eliminating all mixed derivatives in virtue of (\ref{hyper}), we arrive at a defining relation, which has to be fulfilled  identically with
respect to the variables $u,u_{y},u_{x},u_{xx}$, $u_{xxx}$. Comparing the coefficients at $u_{xxx}$ in this relation, we get
\begin{equation}\label{opreq2}
\frac{d}{dy}\frac{\p F}{\p u_{xx}}+3\frac{d}{dx}\frac{\p\Psi }{\p u_x}=0.
\end{equation}
If some equation  from the list (\ref{list1})--(\ref{list14a}) is taken for the $x$-symmetry  then 
the function $F$ is known and the defining relation can also be  splited with respect to $u_{xx}$.

For example, let equation (\ref{list6}) be an $x$-symmetry for
(\ref{hyper}). Then the $u_{xx}$-splitting of (\ref{opreq2}) gives rise to:
$$
\begin{aligned}
&(u_x^2+1)^2\frac{\p^2\Psi }{\p u_x^2}-u_x(u_x^2+1)\frac{\p\Psi }{\p u_x}+(u_x^2-1)\Psi =0,\\
&(u_x^2+1)\left(\Psi\frac{\p^2\Psi }{\p u_x\p u_y}+u_x\frac{\p^2\Psi }{\p u\p u_x}\right)-u_x^2\frac{\p\Psi }{\p u}-u_x\Psi \frac{\p\Psi }{\p u_y}=0.
\end{aligned}
$$
The general solution of this system is given by
\begin{equation*}
\Psi =\sqrt{u_x^2+1}\left(g(u,u_y)+C\ln(u_x+\sqrt{u_x^2+1})\right).
\end{equation*}
Substituting this expression into (\ref{opreq1}) and finding the coefficient at $u_{xx}^3$, we obtain $C=0$
and therefore
\begin{equation}\label{sol1}
\Psi =g(u,u_y)\sqrt{u_x^2+1}.
\end{equation}
Splitting  (\ref{opreq1}) with respect to  $u_{xx}$ and $u_{x},$ we obtain
that (\ref{opreq1}) is equivalent to a system consisting of
(\ref{sol1}) and equations
\begin{equation}\label{eq3}
\begin{aligned}
&g\frac{\p^2g}{\p u\p u_y}-\frac{\p g}{\p u}\frac{\p g}{\p u_y}=0,\qquad  \wp'(u)u_y=2\frac{\p g}{\p u}\frac{\p g}{\p u_y},\\
&g^2\frac{\p^2 g}{\p u_y^2}+g\left(\frac{\p g}{\p u_y}\right)^2-3g \wp+\frac{\p^2 g}{\p
u^2}=0,
\end{aligned}
\end{equation}
where  $ ( \wp')^2=4 \wp^3-g_2 \wp-g_3.$  Since  $\wp'\ne0$ we have $g_u\ne0$ and $g_{u_y}\ne0.$
It follows from the  first two equations (\ref{eq3}) that $g= \sqrt{\wp(u)-\mu}\sqrt{u_y^2+a}$, where $\mu$ and $a$
are constants of integration. The third equation is equivalent to the algebraic equation $4 \mu^3-g_2 \mu-g_3=0$ for
$\mu$. Thus, we get equation (\ref{sh4}).

To prove the theorem we perform similar computations for each equation
from the list (\ref{list1})--(\ref{list14}) taken for $x$-symmetry. For
equations (\ref{list1}), (\ref{list3}) and  (\ref{list7}) the corresponding
hyperbolic equation does not exist. In contrast, equation (\ref{list11}) is an
$x$-symmetry for several different hyperbolic equations. Indeed, in this
case calculating the coefficient at $u_{xx}$ in (\ref{opreq2}), we get
$$
2(u_x+1)^2\frac{\p^2\Psi }{\p u_x^2}-(u_x+1)\frac{\p\Psi }{\p u_x}+\Psi =0,
$$
which implies $\Psi =f_1(u,u_y)(u_x+1)+f_2(u,u_y)\sqrt{u_x+1}$. Substituting this into (\ref{opreq2}), we obtain
$$
\begin{aligned}
&\left(u\frac{\p f_1}{\p u_y}-1\right)\left(u^2f_1\frac{\p f_1}{\p u_y}-3uf_1+2u_y\right)=0,\\
&u^2\left(f_1\frac{\p f_1}{\p u_y}+\frac{\p f_1}{\p u}\right)-2uf_1+2u_y=0,\\
&f_2=2f_1-\frac{2}{u}u_y-uf_1\frac{\p f_1}{\p u_y}.
\end{aligned}
$$
If the first factor in the first equation is equal to zero, we arrive at (\ref{sh14b}). If the second factor equals zero,
then we get
$$
f_1=\frac{2u_y\sqrt{au_y+1}}{u(1+\sqrt{au_y+1})},
$$
where $a$ is a constant. The case $a\ne0$ corresponds to (\ref{sh14}), while $a=0$ leads to equation (\ref{sh14b}) with $c=0$.
The limit $a\to\infty$ gives us equation (\ref{sh14a}).

The computations for remaining $x$-symmetries from the list except for the Swartz-KdV equation (\ref{list8}) are very similar and we do not demonstrate them here. 

Consider the Swartz-KdV equation (\ref{list8}). Equation (\ref{list8}) is exceptional because there is a wide class of hyperbolic equations with this $x$-symmetry. Not all equations from this class are integrable and we derive those of them that have $y$-symmetries.

It is easy to verify that equation
\begin{equation}\label{Swhyp}
u_{xy}=f(u,u_y)u_x
\end{equation}
has the following symmetry
\begin{equation}\label{exeptional}
u_t=u_{xxx}-\frac{3u_{xx}^2}{2u_x}+q(u)u_x^3,
\end{equation}
where
\begin{equation}\label{skdvS1}
\left( \frac{\p }{\p u}+f\frac{\p }{\p u_y}\right)^2f+2qf+q' u_y=0.
\end{equation}
The function $q(u)$ can be eliminated by  an appropriate transformation
$u\to\vph(u)$, but we prefer to use transformations of this type for bringing
the $y$-symmetry to one of equations (\ref{list1})--(\ref{list14}). Here and in the sequel
we make the transformation $x\leftrightarrow y$ in formulas
(\ref{list1})--(\ref{list14}) as well as in other formulas we need.

Any of $y$-symmetries has the form
$$
u_t=u_3+A_2(u,u_1) u_2^2+A_1(u,u_1 )u_2+A_0(u,u_1),\qquad u_n=\frac{\p^nu}{\p u_y^n}.
$$
Equation (\ref{opreq2}) with $x\leftrightarrow y$ is equivalent to
\begin{equation}\label{eqsSK}
\begin{aligned}
&3\frac{\p^2 f}{\p u_y^2}+2 \frac{\p (f A_2)}{\p u_y}+2\frac{\p A_2}{\p u}=0,\\
&3u_y\frac{\p^2 f}{\p u\p u_y}+3f\frac{\p f}{\p u_y}+2A_2u_y\frac{\p f}{\p u}+f\frac{\p A_1}{\p u_y}+2A_2f^2+\frac{\p A_1}{\p u}=0.
\end{aligned}
\end{equation}
Equations  (\ref{skdvS1}) and (\ref{opreq1}) give rise to additional restrictions for the functions $f$ and
$q$.

 For symmetries  (\ref{list1})--(\ref{list4}) we have $A_1=A_2=0$ and equations (\ref{eqsSK})
imply $f=u_y g(u)+h(u)$, $gh=0,\ g'+g^2=0$. In the case $g\ne0$ we get (\ref{sh10a}) with $a=0.$
 For $g=0$ it follows from (\ref{skdvS1}) that $q'=0$ and $f''+2qf=0$.  If $q\ne0$, then without loss of generality we take
$q=-\frac12$ and arrive at equation (\ref{sh10c}).
In the case $g=0, q=0$ we get equation (\ref{sh6}).

For symmetries (\ref{list5}) and (\ref{list6})  $A_1=0, A_2=-3/2\,u_y(u_y^2+1)^{-1}$. It follows from (\ref{skdvS1}) and
(\ref{eqsSK}) that $f=h(u)u_x\sqrt{u_y^2+1}\,, h''=2h(h^2+c_0), q=c_0-3/2\,h^2$.
If $h'=0$, then we put $h=1$ and obtain equation (\ref{sh8}).  In the case $h'\ne0$ we get $h=\sqrt{\wp-\mu}$,
$q=-3/2 \wp$. The hyperbolic equation is given by (\ref{sh4}) with $x\leftrightarrow y$ and $a=0$.

For symmetries (\ref{list7}), (\ref{list8})  $A_2=-\frac32u_y^{-1},\ A_1=0.$ It follows from (\ref{eqsSK}) that $f=g(u)u_y.$
So, we obtain the  equation $u_{xy}=g(u)u_xu_y.$ Both $x$- and $y$-symmetries of the equation take the form (\ref{exeptional}), where
$$
q=C\exp\left(-2\int g(u)\,du\right)-g'-\frac{1}{2} g^2.
$$
The equation is equivalent to the D' Alembert equation  $u_{xy}=0$  under the following transformation:
$$
\bar u=\int du\exp\left(-\int g(u)\,du\right).
$$

For symmetries (\ref{list9}) and (\ref{list10})  $A_2=-\frac34u_y^{-1},\ A_1=0$. It follows from (\ref{eqsSK}), (\ref{skdvS1}) that
$f=g(u)u_y+C\sqrt{u_y},\,\, gC=0,\,\, qC=0,\,\, g'+g^2=0, \,\, q'+2qg=0$. If $C\ne0$, then
$q=g=0$. Taking $C=2$, we get (\ref{sh7}). If $C=0$,
then $g=u^{-1}, q=c_0u^{-2}$, and we obtain (\ref{sh10a}) with $a=0$.

If the $y$-symmetry has the form (\ref{list11}), then it follows from (\ref{skdvS1}), (\ref{eqsSK})
that $f=ku^{-1}(u_y+1-\sqrt{u_y+1})$,\,\, $(k-1)(k-2)=0,\,\, q=3(2-k)/(8u^2)$. If $k=1$ we get (\ref{sh14b}) with  $x\leftrightarrow y$.
The case $k=2$ leads to (\ref{sh14a}).

In the case of $y$-symmetry (\ref{list12}) the system of equations (\ref{skdvS1}), (\ref{eqsSK}) has two
solutions corresponding to equations (\ref{sh11a}), (\ref{sh11c}) with  $x\leftrightarrow y$.

Symmetry (\ref{list13}) gives rise to equation (\ref{sh12}) with  $x\leftrightarrow y$.

Symmetry (\ref{list14}) corresponds to the following equation
$$
u_{xy}=\frac{u_x u_y}{u+a}-(u+a)u_x.
$$
The shift $u\to u-a$ brings it to a special case of equation  (\ref{sh13}).

Considering the linear $x$-symmetry (\ref{list14a}), we obtain equation (\ref{sh10a}) with arbitrary parameter $a$,
equation (\ref{sh10b}), and
\begin{equation}\label{L2}
u_{xy}=a\,u_x+f(u_y-a\,u),
\end{equation}
where $f$ satisfies some nonlinear third order ODE.
The requirement of the  existence of a $y$-symmetry leads to (\ref{shL}).

More detailed information of each equation of the list
(\ref{sh1})--(\ref{sh10b}) can be found in Appendix 1.

\section*{Discussion.}

The hyperbolic equations of the form (\ref{hyper}) having both $x$ and $y$-integrals were described in \cite{zibsok}. In particular, it was shown that any such equation possesses both $x$ and $y$ higher symmetries depending on arbitrary functions. Although not all of these symmetries are integrable,  usually some integrable symmetries exist for such equations.

There are integrable equations having  only $y$-integrals (or only $x$-integrals). An example of such equation is given by (\ref{sh4}), where $a=0$. Namely, the equation
\begin{equation}\label{oneInt}
u_{xy}= \xi'(u) u_y\sqrt{u_x^2+1},
\end{equation}
where $ \xi'(u)=\sqrt{\wp-\mu}$, has the following first order $y$-integral
$$
I=(u_x+\sqrt{u_x^2+1}\,)\, e^{-\xi}
$$
and has no $x$-integrals for the generic Weierstrass function $\wp$. Notice that the same formula gives an $y$-integral for (\ref{oneInt}) with arbitrary function $\xi$.

In some sense equations (\ref{hyper}) having integrals can be reduced to ODEs.
If we are looking for equations (\ref{hyper}) integrable by the inverse scattering transform method, we should concentrate on
integrable equations (\ref{hyper}) without integrals. There are two classes of such equations.
The first one consists of the Klein-Gordon  equation\linebreak $u_{xy}=c u, \, c \ne 0$ and equations related through differential substitutions to the Klein-Gordon
 equation. The symmetries for such equations are  $C$-integrable  (in terminology by F. Calogero).
The second class of hyperbolic integrable equations having no integrals contains equations that can not be reduced to a linear form by  differential substitutions.
This the most interesting class consists of hyperbolic equations admitting only  $S$-integrable higher symmetries.
Such equations can be regarded as $S$-{\it integrable} hyperbolic equations.

For the first glance the anzats (\ref{kdvtype}) seems to be very restrictive if we want to describe all $S$-integrable equations (\ref{hyper}). 
The first question is: why only third order
equations are taken for symmetries? We can justify it in the following way. All known $S$-integrable hierarchies of evolution equations (\ref{sym1})
contain either a third order or a fifth order equation. For polynomial equations this is not an observation but a rigorous statement \cite{sw}.
That is why it is enough to consider hyperbolic equations with symmetries of third order  (sin-Gordon type equations) and hyperbolic equations with
fifth order symmetries (Tzitzeica type equations). The following Tzitzeica type $S$-integrable equations are known \cite{zibsok,Bor1}:
\begin{align}
&u_{xy}=c_1e^u+c_2e^{-2u},  \label{ts1}
\\
&u_{xy}=S(u)f(u_x)g(u_y),   \label{ts3}
\\
&u_{xy}=\frac{\omega'+3 c}{\omega(u)} f(u_x)g(u_y),  \label{shts3a}
\\
&u_{xy}=h(u)\,g(u_y),\ \ h''=0,  \label{ts4}
\end{align}
where
$$
\begin{aligned}
&(f+2u_x)^2(u_x-f)=1,\ \ (g+2u_y)^2(u_y-g)=1,\\
&(S'-2S^2)^2(S'+S^2)=c_1,\ \ \omega '^2=4 \omega^3+c^2.
\end{aligned}
$$
We are planning to consider the Tzitzeica type equations separately. One of the problems here is that the list of  integrable fifth order evolution equations from \cite{mss} possibly is not complete.

The second question is: why we restrict ourselves by symmetries of the form   $u_{t}= u_{xxx}+F(u,u_{x},u_{xx})$ instead of general symmetries of the form
\begin{equation} \label{gen3}
u_{t}= \Phi(u,u_{x},u_{xx},u_{xxx}) \, ?
\end{equation} 
The main reason is the following statement (see \cite{zib}): suppose equation (\ref{gen3}) is a symmetry for equation (\ref{hyper}). Then
$$
\frac{d}{dy} \Big(\frac{\partial \Phi(u,u_x,u_{xx},u_{xxx})}{\partial u_{xxx}}\Big) =0.
$$
Therefore, if we assume that (\ref{hyper}) has no nontrivial integrals, then $$\frac{\partial \Phi(u,u_x,u_{xx},u_{xxx})}{\partial u_{xxx}}=const. $$.

\vskip.3cm \noindent {\bf Acknowledgments.} Authors thank A.V. Zhiber and S.J. Statsev for fruitful
discussions.   V.S. was partially supported by
the RFBR grants 08-01-00440 and NS 3472.2008.2.
A.M. was  supported by Russian Federal Agency for Higher Education. Project 1.5.07.

\section*{Appendix 1. Symmetries, integrals and differential\\ substitutions}
Here an information of the integrable equations from the list  (\ref{sh1})--(\ref{sh10b}) is presented.
Only the simplest equation from the equivalence class is shown. The existence of $x$-integrals
$J(u,u_y,u_{yy},\dots)$ and $y$-integrals $I(u,u_x,u_{xx},\dots)$ was checked till the seventh order.

Linearizing substitutions from Liouville type equations to $v_{xy}=0$ have the form $I=v_x$ or $J=v_y$.
More complicated  substitutions $I=f(v_x,v_{xx},\dots)$ or $J=g(v_y,v_{yy},\dots)$ are presented explicitly.

{\bf Equation (\ref{sh1}).} The symmetries have the following form:
$$
u_t=u_{xxx}-\frac{c}{2}u_x^3-\frac{3}{2}f^2(u)u_{x},\ \ \ u_t=u_{yyy}-\frac{3u_yu_{yy}^2}{2(u_y^2+1)}-\frac{c}{2}u_y^3.
$$

There are two the $sin$-Gordon type equations:

{\bf(\ref{sh1}a).} $u_{xy}=u\sqrt{u_x^2+1}$\,;\qquad {\bf(\ref{sh1}b).}  $u_{xy}=\sin u\sqrt{u_x^2+1}$\\[1ex]
and two Liouville type equations:

{\bf(\ref{sh1}c).} $u_{xy}=\sqrt{u_x^2+1}$ ;  \qquad  the integrals are:
$$
I=\frac{u_{xx}}{\sqrt{u_x^2+1}},\quad J=u_{yy}-u;
$$
the linearizing substitution $u_x=\sinh(y+ v_x)$ gives rise to the general solution:
$$
u=\int\sinh(y+f(x))\,dx+g(y);
$$

{\bf(\ref{sh1}d).}  $u_{xy}=e^u\sqrt{u_x^2+1}$; \qquad  the integrals are:
$$
I=\frac{u_{xx}}{\sqrt{u_x^2+1}}-\sqrt{u_x^2+1},\quad J=u_{yy}-\frac12 u_y^2-\frac12 e^{2u},
$$
The general solution is given by
$$
\begin{aligned}
&u(x,y)=\ln\left(\frac{-\vph(x)g'(y)}{\big(g(y)+h(x)\big)\big(\vph(x)+f(x)\big(g(y)+h(x)\big)\big)}\right),\\[1ex]
&\vph(x)=\exp\left(\int\frac{f(x)}{4f'(x)}\,dx\right),\quad h(x)=\int\frac{f'(x)\vph(x)}{f^2(x)}\,dx.
\end{aligned}
$$

{\bf Equation (\ref{sh2}).} Both $x$- and $y$-symmetries have the  form (\ref{list4}), where $c_1=c_2=0$.
If $ab\ne0$, then  we have the $sin$-Gordon equation.

{\bf(\ref{sh2}a).} $u_{xy}=e^u$ is the Liouville equation. Its symmetries have the same form
as for the $sin$-Gordon equation. The integrals were shown in the Introduction.  The general solution
$$
  u(x,y)=\log{\left( \frac{2 f'(x) g'(y)}{(f(x)+g(y))^2}\right)};
$$ was found by  Liouville in 1853.

{\bf Equation (\ref{sh3}).} The $x$-symmetry has the form  (\ref{list9}), where $c_1=0,c_2=-3/4$;
the $y$-symmetry is of the form (\ref{list5}), where $c_1=c_2=0$. It is an S-integrable equation.

{\bf Equation (\ref{sh4a}).} Both $x$- and $y$-symmetries have the  form (\ref{list5}), where $c_1=0,c_2=-1/2$.
It is an S-integrable equation.

{\bf Equation (\ref{sh4}).} The $x$-symmetry is of the form (\ref{list6}), the form of the $y$-symmetry is analogous:
$$
u_t=u_{yyy}-\frac{3u_yu_{yy}^2}{2(u_y^2+a)}-\frac{3}{2}\wp(u)u_y(u_y^2+a).
$$
If $a=0,$ then this symmetry is  equivalent to  (\ref{list8}).

In the general case the equation can be rewritten using the Jacobi function $\sn$ as:
\begin{equation}\label{Jacob}
u_{xy}=\frac{1}{\sn(u,k)}\sqrt{\vphantom{u_y^2}u_x^2+1}\sqrt{u_y^2+a}.
\end{equation}
This is an S-integrable equation except for the degenerate cases considered below. Notice that the formulas  
$$
\sqrt{\wp(u,g_2,g_3)-\mu_1}=\frac{\cn(u,k)}{\sn(u,k)},\quad \sqrt{\wp(u,g_2,g_3)-\mu_2}=\frac{\cn(u, k)}{\dn(u, k)}.
$$
lead to another forms of this equation. They  can be reduced to (\ref{Jacob}) by the substitution $(u,k)\to(\lambda u,f(k))$ (see \cite{Batmen}, Sec. 13.22).

There are two degenerate cases for the Weierstrass function. In the first case when $\wp(u)=u^{-2}$ 
we have $\mu=0$ and $\sqrt{\wp-\mu}=u^{-1}$. In the second case $\wp(u)=\sin^{-2}u-\frac13$, $\mu=-\frac13$ and $\sqrt{\wp-\mu}=\sin^{-1} u$.

{\bf(\ref{sh4}a).} Equation $u_{xy}=u^{-1}\sqrt{u_x^2+1}\sqrt{u_y^2+a}$ \ is C-integrable, the integrals are:
$$
I=\frac{u_{xx}}{\sqrt{u_x^2+1}}+\frac1u\sqrt{u_x^2+1},\quad J=\frac{u_{yy}}{\sqrt{u_y^2+a}}+\frac1u\sqrt{u_y^2+a}.
$$
The general solution is given by:
$$
u(x,y)=\sqrt{f(x)+g(y) }\,\left(-\int\frac{dx}{f'(x)}-a\int\frac{dx}{g'(y)}\right)^{1/2}.
$$

{\bf(\ref{sh4}b).} Equation $u_{xy}=(\sin u)^{-1}\sqrt{u_x^2+1}\sqrt{u_y^2+a}$ \ is C-integrable, the integrals are:
$$
I=\frac{u_{xx}}{\sqrt{u_x^2+1}}+\cot u\sqrt{u_x^2+1},\quad J=\frac{u_{yy}}{\sqrt{u_y^2+a}}+\cot u\sqrt{u_y^2+a}.
$$
If $a=0,$ then the general solution is
$$
u(x,y)=2\arccos\left(\frac{f(x)+h(x)+g(y)}{2f(x)}\right)^{1/2},\quad h(x)=\int\sqrt{f'^2-f^2}\,dx.
$$
If $a\ne0,$ then the general solution is
$$
\begin{aligned}
&u(x,y)=\arccos\Psi (x,y), \\
&\Psi (x,y)=\frac{1}{2}w(x)\left[e^g(\xi+h)^2-e^{-g}\right](2w'+fw)+(\xi+h)e^g,\quad g=g(y),  \\
&h(y)=\int e^{-g}\sqrt{g'^2-a}\,dy,\quad f'(x)=\frac{1}{2}(1+f^2)-2\frac{w''}{w},\quad \xi(x)=\int\frac{dx}{w^2(x)}.
\end{aligned}
$$

{\bf(\ref{sh4}c).} $a=0,\ u_{xy}=f(u) u_y\sqrt{u_x^2+1}$. There exists the following $y$-integral

$$
I=(u_x+\sqrt{u_x^2+1})\exp(-\xi(u)), \quad \xi(u)=\int f(u)\,du
$$
for all $f(u)$. This leads to  the first order ODE:
$$
u_x=\dfrac12\left(h(x)e^\xi-\big(h(x)e^\xi\big)^{-1}\right).
$$

All remaining equations  are C-integrable. Some of them have two integrals and can be
reduced to  the D'Alembert  equation. Others have no integrals and can be reduced to  the Klein-Gordon  equation.

{\bf Equation (\ref{sh6}).} The $x$-symmetry has the form (\ref{list8})  and
the $y$-symmetry is the mKdV equation $u_t=u_{yyy}-6u^2u_y$. The integrals are:
$$
I=\frac{u_{xxx}}{u_x}-\frac{3u_{xx}^2}{2u_x^2},\quad J=u_y-u^2.
$$
The general solution is given by
$$
u(x,y)=\frac{g''(y)}{2g'(y)}-\frac{g'(y)}{f(x)+g(y)}.
$$

{\bf Equation (\ref{sh7}).} The $x$-symmetry has the form (\ref{list8})  and
the $y$-symmetry is (\ref{list9}), where $c_1=0,c_2=-3$. The integrals are:
$$
I=\frac{u_{xxx}}{u_x}-\frac{3u_{xx}^2}{2u_x^2},\quad J=\sqrt{u_y}-u.
$$
The general solution is given by
$$
u(x,y)=-\frac{g'(y)}{f(x)+g(y)}+\int\frac{(g'')^2}{4g'^2}\,dy.
$$

{\bf Equation (\ref{sh8}).} The $y$-symmetry takes the form (\ref{list5}),  where $c_1=0,c_2=-1/2$  and
the $x$-symmetry is
$$
u_t=u_{xxx}-\frac{3u_{xx}^2}{2u_x}-\frac{1}{2}u_x^3.
$$
This symmetry can be reduced to  (\ref{list8}) by $u\to \ln u$. The integrals are:
$$
I=\frac{u_{xxx}}{u_x}-\frac{3u_{xx}^2}{2u_x^2}-\frac{1}{2}u_x^2,\quad J=(u_y+\sqrt{u_y^2+1}\,)e^{-u}.
$$
The general solution is given by
$$
u(x,y)=\ln\left[1+\frac{g(y)}{f(x)+h(y)}\right]+\int g^{-1}\sqrt{g'^2-g^2}\,dy,\quad h=-\frac{1}{2}g-\frac{1}{2}\int\sqrt{g'^2-g^2}\,dy.
$$

{\bf Equation (\ref{sh10}).}(The Goursat equation.) Both $x$- and $y$-symmetries  have the form (\ref{list10}) with arbitrary  constant $c$.

The equation is reduced to the Klein-Gordon equation $v_{xy}=\frac{1}{4}v$ by any  of the following two differential
substitutions:
$$
(1)\ u_x=4v_x^2,\quad u_y=v^2;\qquad (2)\ u_x= v^2,\quad u_y=4v_y^2.
$$

{\bf Equation (\ref{sh10a}).}  The $x$-symmetry has the form (\ref{list10}),  where $c=0$  and
the $y$-symmetry can be obtained from (\ref{list4}) by the substitution $c_2=0,\ u\to -\ln u$. Moreover, there exists 
the following second order $y$-symmetry $u_t=u_{yy}-2u^{-1}(u_y^2+au_y)$.

The integrals and the general solution are:
$$
I=\frac{u_{xx}}{u_x},\quad J=\frac{u_y+a}{u};\quad u(x,y)=\frac{f(x)-ag(y)}{g'(y)}.
$$

{\bf Equation (\ref{sh10c}).}  The $x$-symmetry has the form (\ref{exeptional}),  where $q=-\frac12$  and
the $y$-symmetry is given by (\ref{list4}),  where $c_1=-\frac32a^2, c_2=-\frac32b^2$.
The integrals are:
$$
I=\frac{u_{xxx}}{u_x}-\frac{3u_{xx}^2}{2u_x^2}-\frac{1}{2}u_x^2,\quad J=u_y-ae^u+be^{-u}.
$$
In the case $a\ne0$ the general solution is given by
$$
u(x,y)=\ln g(y)+\ln\left[1+\frac{h(y)}{f(x)-a\vph(y)}\right],\quad \ln h=\int (ag+bg^{-1})\,dy,\quad \vph=\int gh\,dy;
$$
if $a=0$ then
$$
u(x,y)=\ln \frac{f(x)-bg(y)}{g'(y)}.
$$

{\bf Equation (\ref{sh11a}).}  The $x$-symmetry has the form  (\ref{list12}). There are the following 
$y$-symmetries:
$$
\begin{aligned}
&u_t=u_{yyy}-\frac{3}{2}(3+\coth u)u_yu_{yy}+\frac{1}{4}(3\coth^2 u+6\coth u+7)u_y^3,\\
&u_t=u_{yy}-\frac 1 2(3+\coth u)u_y^2.
\end{aligned}
$$
The integrals are:
$$
I=\frac{e^{-u}\eta^2-2\eta+e^u}{\sinh u},\quad J=\frac{u_{yy}}{u_y}-\frac12u_y(\coth u+3).
$$
The general solution is given by:
$$
u(x,y)=-\frac{1}{2}\ln(1+\psi^2),\quad \psi=f(x)(g(y)+h(x)),\quad f'=f-\frac{1}{4}f^3h'^2.
$$

{\bf Equation (\ref{sh11c}).} The $x$-symmetry has the form  (\ref{list12}). There are the following 
$y$-symmetries:
$$
u_t=u_{yyy}-6u_yu_{yy}\coth u+2(3\coth^2 u-1)u_y^3,\quad u_t=u_{yy}-2u_y^2\coth u.
$$
The integrals are:
$$
I=\frac{\eta-e^u}{\eta-e^{-u}},\quad J=\frac{u_{yy}}{u_y}-2u_y\coth u.
$$
The general solution is:
$$
u(x,y)=\frac{1}{2}\ln\left|\frac{\psi+1}{\psi-1}\right|,\quad \psi=f(x)(g(y)+h(x)),\quad h'=-\frac{f'^2+4f^2}{4f^3}.
$$

{\bf Equation (\ref{sh11b}).} Both $x$- and $y$-symmetries have the form (\ref{list12}).
The equation is reduced to the Klein-Gordon one $v_{xy}=v$ by  the following differential
substitution:
$$
u_x=\big(v^{-1}v_x\sinh u+\cosh u\big)^2-1,\quad u_y=\big( v^{-1}v_y\sinh u+\cosh u\big)^2-1.
$$

{\bf Equation (\ref{sh14b}).} There are $x$-symmetry of the form  (\ref{list11}) and
the following $y$-symmetry:
$$
u_t=u_{yyy}-\frac{3u_yu_{yy}}{2u}+\frac{3u_y^3}{4u^2}-\frac{3c}{4}(2uu_{yy}+2u_y^2-cu^2u_y).
$$
The equation can be reduced to the Klein-Gordon equation $v_{xy}=cv$ by  the following differential
substitution:
$$
u=v^2/z,\quad z_x=-v_x^2,\ z_y=-cv^2.
$$
If $c=0$ then the Klein-Gordon equation is reduced to the D'Alembert equation and the following two integrals appear:
$$
I=\frac{(\eta-1)^2}{u},\quad J=\frac{u_{yy}}{u_y}-\frac{u_y}{2u}.
$$
The general solution is:
$$
u(x,y)=\frac{(f(x)+g(y))^2}{z(x)},\quad z(x)=-\int f'^2(x)\,dx .
$$
Notice that if $c=0$ the equation admits a second order symmetry.

{\bf Equation (\ref{sh14a}).} There are $x$-symmetry of the form  (\ref{list11}) and two
the following $y$-symmetries:
$$
 u_t=u_{yyy}-6u^{-1}u_yu_{yy}+6u^{-2}u_y^3,\quad u_t=u_{yy}-2u^{-1}u_y^2.
$$
The integrals and the general solution are given by:
$$
I=\frac{\eta-1}{u},\quad J=\frac{u_{yy}}{u_y}-2\frac{u_y}{u};\quad
u(x,y)=\frac{f^2(x)}{h(x)+g(y)},\quad h(x)=-\int f'^2(x)\,dx.
$$

{\bf Equation (\ref{sh14}).} Both $x$- and $y$-symmetries  have the form (\ref{list11}).
The integrals are of the form:
$$
I=\frac{u_{xx}}{\eta(\eta -1)}-\frac{2}{u}\eta(\eta -1),\quad J=\frac{u_{yy}}{\xi (\xi -1)}-\frac{2}{u}\xi (\xi -1).
$$
The general solution is given by:
$$
 u(x,y)=\frac{(f(x)+g(y))^2}{z(x,y)},\quad z(x,y)=-\int f'^2(x)\,dx-\int g'^2(y)\,dy.
$$

{\bf Equation (\ref{sh12}).} There are $x$-symmetry of the form  (\ref{list13}) and two
the following $y$-symmetries:
$$
u_t=u_{yyy}-9u^{-1}u_yu_{yy}+12u^{-2}u_y^3,\quad u_t=u_{yy}-3u^{-1}u_y^2.
$$
The integrals are of the form:
$$
I=\frac{u_x}{u}+u^2,\quad J=\frac{u_{yy}}{u_y}-3\frac{u_y}{u}.
$$
The  general solution is:
$$
u(x,y)=\left(\frac{f'(x)}{2\big(f(x)+g(y)\big)}\right)^{1/2}.
$$

{\bf Equation (\ref{sh13}).} There are $x$-symmetry of the form  (\ref{list14}) and two
the following $y$-symmetries:
$$
u_t=u_{yyy}-3u^{-1}(2u_y+a)u_{yy}+3au^{-2}u_y(3u_y+a)+6u^{-2}u_y^3,\quad u_t=u_{yy}-2u^{-1}u_y(u_y+a).
$$
When $a=0$ the $y$-symmetry (\ref{list8}) is also admitted.
The equation can be reduced to the Klein-Gordon one $v_{xy}=-av$ by the following substitution:
$$
u_x=\left(\frac{v_x}{v}-u\right)(u-\lambda ),\quad u_y=\frac{1}{\lambda }\left(u\frac{v_y}{v}+a\right)(u-\lambda),
$$
where $\lambda$ is arbitrary parameter.
If $a=0,$ then the Klein-Gordon equation is reduced to the D'Alembert equation 
and two integrals appear:
$$
I=\frac{u_x}{u}+u,\quad J=\frac{u_{yy}}{u_y}-2\frac{u_y}{u},
$$
In this case the  general solution is in the form $u(x,y)=f'(x)(f(x)+g(y))^{-1}$.

{\bf Equation (\ref{shL}).}  The $x$-symmetry is $u_t=u_{xxx}-\frac32\,a\,u_{xx}$
and the $y$-symmetry has the form  (\ref{list10}), where $c=0$ and $x\to y$.
The integrals are of the form:
$$
I=u_{xxx}-\frac{3}{2}\,a\,u_{xx}+\frac{a^2}{2}\,u_x,\quad J=\frac{u_{yy}}{a\,u_y+\sqrt{u_y}}.
$$
The  general solution is given by:
$$
u(x,y)=f(x)+e^{ax}\int \left(g(y)+\frac{1-e^{-ax/2}}{a}\right)dy.
$$
The limit $a\to 0$ is available here.

{\bf Equation (\ref{sh10b}).} There are infinitely many symmetries of the
form $u_t=P(\p_x,\p_y)u,$ where $P$ is an arbitrary polynomial with constant coefficients. In particular, 
there exist  $x$- and $y$-symmetries of the form $u_t=P_1(\p_x)u$ and $u_t=P_2(\p_y)u$ .
If $c\ne 0$ integrals do not exist otherwise the simplest integrals are:
$I=u_x,\,J=u_y$.

\end{document}